\documentstyle[emulapj]{article}

\font\tenbg=cmmib10 at 10pt

\epsfverbosetrue

\font\tenbg=cmmib10 at 10pt

\def \rvectheta{{\hbox{\tenbg\char'022}}}

\begin{document}

\lefthead{MAGNETIC FIELD LIMITATIONS}

\righthead{BISNOVATYI-KOGAN AND LOVELACE}

\submitted{}

\title{Magnetic Field Limitations on
Advection Dominated Flows}

\medskip

\author{G.S. Bisnovatyi-Kogan}
\affil{Space Research Institute,
Russian Academy of Sciences,
Moscow, Russia; gkogan@mx.iki.rssi.ru}

\author{R.V.E. Lovelace}
\affil{Department of Astronomy,
Cornell University, Ithaca, NY 14853-6801;
rvl1@cornell.edu}

\slugcomment{Submitted to the Astrophysical Journal}

\begin{abstract}

  Recent papers
discussing advection dominated accretion flows
(ADAF) as a solution for
astrophysical accretion
problems should be treated with
some caution because of their
uncertain physical basis.
  The suggestions underlying
ADAF involve ignoring the
magnetic field reconnection
in heating of the plasma flow,
assuming electron heating due {\it only} to
binary Coulomb collisions with ions.
   Here, we analyze the
physical processes in optically
thin accretion flows at low
accretion rates including the
influence of an equipartition
random magnetic field and heating
of electrons due to magnetic
field reconnection.
  The important role of the magnetic
field pointed out by Shwartsman (1971)
comes about because the magnetic energy density
$E_m$ increases more rapidly with
decreasing distance than the
kinetic energy density $E_k$ (or thermal
energy density).
  Once $E_m$ grows to a
value of order $E_k$,
further accretion to smaller distances  is
possible {\it only} if magnetic flux
 is destroyed
by reconnection.
 For the smaller distances it is likely that
there is approximate
equipartition
$E_m \approx E_k$.
  Associated with the destruction of magnetic
flux there is  dissipation of
 magnetic energy.
  We argue that
the field annihilation leads to
appreciable electron heating.
Such heating significantly restricts the
applicability of ADAF solutions.
  Namely, the radiative efficiency of
the flows cannot be less
than about $25\%$ of the standard
accretion disk value.
 This result casts doubt on
attempts to connect ADAF models
with the existence of
event horizons of  black holes,
and to explain the low
luminosities of
massive black holes in
nearby galactic nuclei.

\end{abstract}

\keywords{accretion,
accretion disks---galaxies:
active---plasmas---magnetic
fields---stars: magnetic
fields---X-rays: stars}

\section{Introduction}
   Recent papers (see Narayan, Barret,
McClintock 1996; Menou, Quataert,
Narayan 1997; Narayan, Mahadevan,
\& Quataert 1999; and references
therein;  see also Ichimarau 1977)
considering  advection dominated
accretion flows
(ADAF) as a solution for many
astrophysical problems should be treated with
some caution, because of its
uncertain physical basis.
  The suggestions underlying ADAF include the
neglect of the heating of the
accretion flow due to magnetic field annihilation
and the assumption of electron
heating due {\it only}  to
binary collisions with protons (ions).
    These issues were first pointed out by
Bisnovatyi-Kogan
and Lovelace (1997; hereafter BKL) and have
subsequently been discussed
further by Quataert (1998), Blackman  (1998), and
others.

\begin{figure*}[t]
\epsscale{1.0}
\plotone{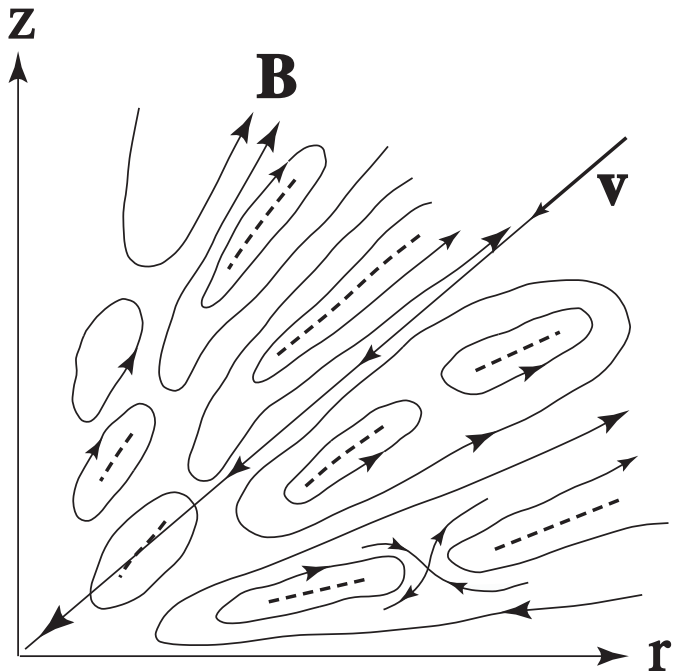}
\caption{
Sketch of the
poloidal projections of
instantaneous
magnetic field lines in a quasi-spherical
accretion flow.  Only one quadrant
of the flow is shown.
    The dashed lines
indicate neutral surfaces where
reconnection or annihilation of
the magnetic field occurs.
 The radial line labeled by ${\bf v}$
shows the time-averaged flow velocity.
}
\label{Figure 1}
\end{figure*}

   The work by Quataert (1998) envisions
an {\it almost uniform} magnetic field
in the accretion flow
which makes
the theory of Alfv\'enic turbulence of
Goldreich and Sridhar (1995) applicable.
   The linear damping of short wavelength
Alfv\'en modes is then calculated to
determine the relative importance of heating
of electrons and ions.
   We believe that this picture
is {\it inconsistent}, because the magnetic
field is strongly non-uniform, in fact,
turbulent with $\delta B/B \sim 1$, and
magnetic flux is necessarily destroyed
in the flow.
   We have a  fundamentally different
picture of  MHD turbulence in accretion
flows (BKL),
sketched in Figure 1, which
shows that there are
many neutral layers
of the magnetic field with
associated current sheets where field
reconnection and annihilation occurs.
  At these neutral layers, there is sporadic
Ohmic dissipation of magnetic energy.
   This picture is qualitatively
similar to that for the
chaotic magnetic field of the corona of
the Sun which we discuss later.

   Competition between
rapid accumulation of observational
data, mainly
from Hubble Space Telescope,
X-ray satellites, and the rapid development
of theoretical models
creates a situation
where a model is sometimes disproved
during the time of its publication.
 Some  ADAF models give examples
of this.
 One is connected with an
explanation of the unusual
spectrum of the galaxy  NGC 4258 which
has a nonmonotonic dependence.
    It was claimed
that such spectrum may be  explained
by an ADAF model (Lasota {\it et al.} 1996).
  However, recent observations
(Herrnstein 1997) give new data
in the X-ray from RXTE (Cannizzo {\it et al.} 1998)
which shows features not explained by ADAF.
  High-resolution radio continuum observations
of this object (Herrnstein {\it et al.} 1998)
put severe constraints on the ADAF model,
indicated that it does not apply inside
$\sim 100$ Schwarzschild radii.

 Another example is connected
with an attempt to prove the existence of
the event horizons of the
black holes in  ADAF models.
 Figure 7
of the work by Menou, Quataert
and Narayan (1997), (or figure 2 from
Narayan, Garcia, and McClintock 1997)
was presented as a proof of the
existence of the event horizon
of black holes and at the same time
as a triumph of the ADAF model.
   Unfortunately, the data for this figure
appears to be incomplete,
and a full set of observational data
smears this picture (Chen {\it et al.} 1997).

   Radio observations of low luminosity cores
of elliptical galaxies, where ADAF models
could explain the X-ray luminosity, put
severe constraints on the
models (Di Matteo {\it et al.} 1998).
The observed radio flux is much smaller than
predicted by the ADAF model with energy
equipartition magnetic fields, and
require magnetic energy density much
below equipartition which is implausible.

   It is of course difficult to
justify a physical model by astronomical
observations without a firm physical basis.
  Here, we analyze the different
processes in an optically
thin accretion flow at low accretion rates.
   Of particular importance is the inclusion of
the influence of a chaotic magnetic field
which is likely embedded in the accreting plasma.
 We show that accurate
account of these processes strongly
restricts the applicability of  ADAF
solutions.
   Namely, the radiative  efficiency of
the accretion flows cannot become less
than about $25\%$ of the standard
accretion disk value.
   This casts doubt on
the attempts to connect ADAF models
 with the existence of the
event horizons of a black holes.
 Also, it makes it unlikely that  ADAF models
explain the low luminosity of
massive black holes in
nearby galactic nuclei.

   In \S 2 we discuss the basic equations
and give a solution for the time-averaged
magnetic field in a quasi-spherical
accretion flow.
  In \S 3 we analyze the energy
dissipation in accretion flows
where there is equipartition
between magnetic energy and flow
energy.
    In \S 4 we point out the
relevance of observations of
magnetic energy dissipation in
the solar corona to that in
accretion flows.
     Conclusions of this work
are given in \S 5.

\section{Magnetic Field Enhancement
in Quasi-spherical Accretion}

 Matter flowing into a black
hole from a companion star or from the
interstellar medium is
likely to be  magnetized.
   Due to the more rapid increase
of the magnetic energy density  in comparison
with kinetic energy densities, the
dynamical influence of
the magnetic field becomes more
and more important as the matter
flows inwards.
   Shwartsman (1971) argued
that beginning even at large radii there
should be
approximate {\it equipartition}
of magnetic and kinetic energies.
   Of course, the main energy
release in an accretion flow
occurs in the region of small radii.
  This equipartition is
usually accepted in recent ADAF models
of accretion (see e.g Narayan and Yi
1994).

  Shwartsman (1971) considered
an averaged, quasi-stationary picture
of accretion of magnetized matter
with local equipartition.
 Another
variant of magnetic accretion
considered by
Bisnovatyi-Kogan and Sunyaev (1972)
(see also Chang and Ostriker 1985)
 is where
equipartition and magnetic field
annihilation is accompanied by formation
of shock waves.
  A more accurate account of the heating of
matter by magnetic field
annihilation was done by Bisnovatyi-Kogan and
Ruzmaikin (1974) where exact
nonstationary solutions for the field
amplification in a spherical
accretion flow were obtained.

  Consider the  general
problem of magnetic field amplification in
a plasma accretion flow.
  The flow is described by the
magnetohydrodynamics (MHD) equations,
$$
\rho { d {\bf v} \over dt}=
-{\bf \nabla} P +\rho {\bf g}
+{1\over c} {\bf J \times B} +
\eta \nabla^2 {\bf v}~,
$$
$$
{\bf \nabla \times B} =
{4 \pi \over c} {\bf J}~,\quad
{\bf \nabla \times E } =
-\frac{1}{c}\frac{\partial {\bf B}}{\partial t}~,
\quad{\bf \nabla \cdot B} = 0~,
$$
\begin{equation}
{\bf J} =\sigma \left({\bf E}
+ {\bf v \times B}/c\right)~,
\end{equation}
where ${\bf v}$ is the flow velocity,
${\bf B}$ the magnetic field,
$P$  the plasma pressure,
$\sigma$  the plasma
conductivity, $\eta$ the dynamic
viscosity (with $\nu=\eta/\rho$
the kinematic viscosity), ${\bf J}$  the
current density, and ${\bf E}$
the electric field.
  These equations can be combined
to give the induction equation,
$$
\frac{\partial {\bf B}}{\partial t}=
{\bf \nabla \times (v \times B) }-
{\bf \nabla \times}(\eta_m {\nabla \times} {\bf B})~,
$$
\begin{equation}
\approx {\bf \nabla \times (v \times B) }+
\eta_m{\bf \nabla}^2 {\bf B}~,
\end{equation}
where $\eta_m \equiv c^2/(4\pi \sigma)$ is the
magnetic diffusivity, and the
approximation involves neglecting
${\bf \nabla}\eta_m$.

  In equations (1) and (2) both the
viscosity $\nu$ and the magnetic
diffusivity $\eta$ have the same
units and both are assumed to be due to turbulence
in the accretion flow.
  Thus, it is reasonable to express
both transport coefficients using
the ``alpha'' prescription of
Shakura (1972) and
Shakura and Sunyaev (1973),
\begin{equation}
\nu = \alpha~ c_s ~\ell_t~,
\quad \quad \eta_m  = \alpha_m ~c_s~ \ell_t~,
\end{equation}
where $\ell_t$ is the outer scale of
the turbulence, and $c_s=\sqrt{P/\rho}$
is the isothermal sound speed.
   Bisnovatyi-Kogan
and Ruzmaikin (1976) introduced
$\alpha_m$ and proposed that
$\alpha_m \sim \alpha$.
   Note that $\alpha$ and $\alpha_m$
characterize a turbulent MHD flow
in which there is a Kolmogorov cascade of
energy from large scales ($\ell_t$)
to much smaller scales where the
actual (microscopic) dissipation
of energy occurs.

  Consider for the moment a
simpler case, that  of
an axisymmetric
poloidal magnetic field,
${\bf B} = B_r \hat{\bf r} +
B_\theta \hat{ \rvectheta~}$.
   Neglect of the toroidal magnetic field
is plausible because this component
of the field increases with
decreasing $r$ much less rapidly than
the poloidal field. (For a perfectly conducting
plasma $B_p \propto 1/r^2$ whereas $B_\phi
\propto 1/\sqrt{r}$.)
For these conditions
we have
$$
B_r = {1\over r^2\sin\theta}
{\partial \Psi \over \partial \theta}~, \quad
B_\theta = - {1\over r \sin \theta}
{\partial \Psi \over \partial r}~,
$$
where $\Psi \equiv r \sin \theta A_\phi$
is the flux function and ${\bf A} $
is the vector potential.
    For quasi-spherical accretion, ${\bf v} \approx
\bar{v}_r\hat {\bf r}$  where $\bar{v}_r <0$
is the time-averaged radial flow velocity.
    Consequently,  Ohm's law in (1)
with $E_\phi =
-(1/c)\partial A_\phi/\partial t$ gives
\begin{equation}
{\partial \Psi \over \partial t}=
-~\bar{v}_r {\partial \Psi \over \partial r} +
\eta_m \Delta^* \Psi~,
\end{equation}
without approximation, where
$$
 \Delta^* \equiv {\partial^2 \over \partial r^2}
+(1-\mu^2){\partial^2 \over \partial \mu^2}
$$
is the adjoint Laplacian operator, and
$\mu \equiv \cos\theta$.

\begin{figure*}[t]
\epsscale{1.2}
\plotone{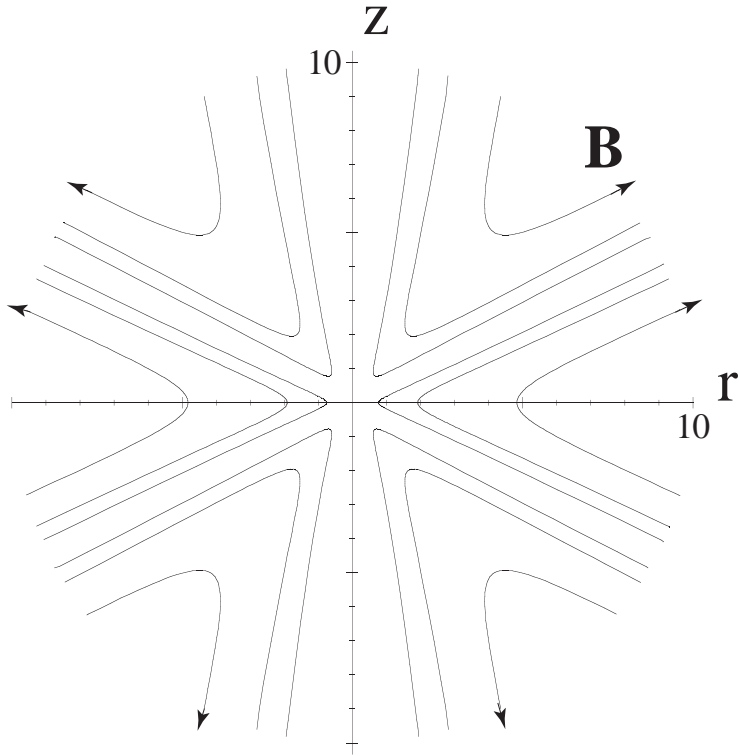}
\caption{
Poloidal projection of
{\it time-averaged}
 magnetic field lines
in quasi-spherical accretion
for conditions where
there is equipartition
between the energy density in the
field and that in the flow.
   The field
lines correspond to
$\Psi =(0.5,1,2)\times$ const, with
$\Psi(r,\theta)$ given by
equation (4) with
$\partial \Psi/\partial t =0$,
and $n=4$ in equation (6).
}
\label{Figure 2}
\end{figure*}

   As a further simplification,
consider stationary
solutions of equation (4).
(Figure 1 shows the instantaneous
configuration.)
  Because $\bar{v}_r \propto c_s \propto v_K
\equiv \sqrt{GM/r}$ and because it is plausible
to assume $\ell_t \propto r$,
equation (4) with $\partial \Psi/\partial t=0$
is homogeneous in $r$ with solutions
of the form
\begin{equation}
\Psi = r^\delta \Theta(\theta)~,
\end{equation}
where $\delta=$const.
 We find that $\Theta$ obeys the equation
\begin{equation}
(1-\mu^2){\partial^2\Theta \over \partial \mu^2}
+n(n-1) \Theta =0~,
\end{equation}
where $\mu \equiv \cos\theta$, and
$n=2,3,..$.
 This is the equation for the Gegenbauer
polynomial $G_n^{(-{1/2})}(\mu)$, where
\begin{equation}
 n(n-1)=\delta \left({r|\bar{v}_r| \over \eta_m}\right)
+\delta(\delta-1)~,
\end{equation}
 and $G_n^{(-{1/2})}(\mu) \propto
(1-\mu^2)d P_{n-1}/d\mu$, where $P_n$ is a
Legendre polynomial.

   For the limit of a highly conducting plasma,
the magnetic field is frozen into the
flow and
$|B_r| \propto r^{-2}$.
  Thus the magnetic
energy-density varies as
${E}_{m} = {\bf B}^2/8\pi \propto r^{-4}$.
On the other hand, as noted
by Shwartsman (1971), the kinetic energy-density
varies as $ E_{k} =
\rho{\bf v}^2/2  \propto r^{-{5/ 2}}$.
    The  magnetic field can
increase only up to the
equipartition value where the magnetic and
kinetic energy-densities  are approximately
equal.
   This occurs at a
large distance $r=r_{equi} \gg r_S$ and
that this is maintained
for smaller $r$, where $r_S$ is
the Schwarzschild radius of the central object
(Shwartsman 1971).
   Further accretion for $r<r_{equi}$ is
possible {\it only} if magnetic flux is destroyed
by reconnection and the magnetic
energy ${ E}_{m}$ is dissipated.
   For these conditions ${E}_{m} \propto
r^{-{5/ 2}}$ so that $|{\bf B}| \propto
r^{-{5/ 4}}$.  In view of equation (5),
this corresponds to $\delta = 3/4$.  Figure 2 shows
the poloidal projection of the field
lines for a sample case with $\delta = 3/4$.

\section{Stationary Magnetic Field Distribution in
Quasi-spherical Accretion}

   In  optically thin accretion
disks at low accretion rates, the density
of the matter is low and energy
exchange between electrons and ions
due to binary collisions is slow.
   In this situation, due to different
mechanisms of heating and
cooling for electrons and ions, the
electrons and ions may
have different temperatures.
  This  was first discussed by Shapiro,
Lightman, and Eardley (1976), but
advection was not included.
  Narayan and Yi (1995) pointed out
that advection in this case is
extremely important.
   It may carry the main
energy flux into a black
hole, giving a low radiative
efficiency of the accretion of $10^{-3}$ to
$10^{-4}$, the ADAF solutions.

   The ADAF solutions assume
an ``alpha'' prescription for the
viscosity.
    Thus, as the ion temperature
increases, the viscosity increases
and the viscous dissipation also
increases.
   If the energy losses by ions
are small, then there can be
a kind of ``thermo-viscous''
instability where heating
increases the viscosity and the
increase in viscosity increases
the heating.
    Development of this instability
leads to formation of ADAF.
  In the ADAF solution the ion
temperature is of the order of the
virial temperature
$\sim GMm_i/r$.
   This means
that even for high initial angular
momentum, the disk is
very thick, forming
a quasi-spherical accretion flow.

   The conclusion about
low radiative efficiency
of ADAF is valid only
when the influence of
a magnetic field is neglected.
   The ${\bf B}$ field is
likely to be turbulent and
in approximate equipartition
with the flow.
   For such conditions there
must be continuous
destruction of magnetic
flux by field reconnection and
annihilation (Shwartsman 1971).
   This reconnnection is
expected to significantly heat
electrons which can efficiently
emit magneto-bremstrahlung
radiation.  Because of the electron
radiation, the radiative
efficiency is {\it not} small.

   As mentioned, the poloidal
magnetic field energy density tends to grow
locally as $E_m^0 \propto r^{-4}$ with
decreasing distance $r$.
   However, as
a result of equipartition
$E_m \propto r^{-5/2}$.
  The difference in the radial gradients
of $E^0_m$ and $E_m$ must therefore go into
Ohmic heating of the
matter.
  This leads to the
important result
\begin{equation}
T\left(\frac{dS}{dr}\right)_{\rm Ohm} =
-\frac{3}{2}~\frac{B^2}{8\pi r \rho }~,
\end{equation}
with $S$ the entropy per unit mass,
 $T$ measured in energy units, and
$B \equiv <{\bf B}^2>^{1/2}$,
where the angle brackets indicate a time
average over the turbulence
(Bisnovatyi-Kogan \& Ruzmaikin 1974).

   In  quasi-spherical
accretion flows, equipartition between
magnetic and kinetic energy of the flow
was proposed by Shwartsman (1971),
\begin{equation}
\frac{B^2}{8\pi} \sim {1\over 2}~\rho~ v_r^2=
\frac{~\rho~ GM}{r}~,
\end{equation}
which corresponds to an
Alfv\'en speed $v_A
\equiv B/\sqrt{4\pi \rho} \sim \sqrt{2}v_K$,
with $v_K\equiv \sqrt{GM/r}$ the Keplerian
speed.
   For disk accretion, where
there is more time for a field dissipation,
Shakura (1972) proposed
 equipartition
between magnetic and
turbulent energy.
   For an ``alpha'' prescription
for the viscosity, where the turbulent
velocity is $v_t = \alpha c_s$,
this
leads to the relation
\begin{equation}
\frac{B^2}{8\pi}  \sim {1\over 2}~\rho~ v_t^2 =
\frac{1}{2}~\alpha^2 P~,
\end{equation}
   In ADAF solutions, where the ion
temperature is of the order of the
virial temperature, the
two expressions, (9)
and (10),  give comparable estimates
for $B^2/8\pi$ if $\alpha \sim 1$.
   Note however that in the following
the exact value of $v_A/v_K$ is not
assumed.

   For quasi-spherical
accretion, the local electric
field strength can be
estimated as ${\cal E} \sim {v_t B}/{c}$.
  The heating of the matter
by  Ohmic dissipation
is then given by
\begin{equation}
T\left(\frac{dS}{dr}\right)_{\rm Ohm} =
-~\frac{ ~\sigma ~{\cal E}^2}{\rho~ v_r}
\sim - ~\frac{~\sigma ~v_t^2 B^2}{\rho~ v_r c^2}~.
\end{equation}
Using equations (3) and $v_t=\alpha c_s$,
we find that equation (11)
is compatible with (8)
if $ \alpha^2/\alpha_m \sim
(3/4)(\ell_t/r)(|v_r|/c_s)$.

  Equations for the radial
variations of the ion
and electron temperatures
can be written as
\begin{equation}
{dE_i\over dt}-{P_i\over\rho^2}{d\rho\over dt}=
{\cal H}_{\eta i}+(1-g){\cal H}_{B}-Q_{ie}~,
\end{equation}
\begin{equation}
{dE_e\over dt}-{P_e\over\rho^2}{d\rho\over dt}=
{\cal H}_{\eta e}+g{\cal H}_{B}+
Q_{ie}-{\cal C}_{brem}-{\cal C}_{cyc}~,
\end{equation}
which differ from  the corresponding equations
of BKL by separating out the viscous and Ohmic
heating contributions.
Here, ${d}/{dt}={\partial}/{\partial t}+
v_r{\partial}/{\partial r}$, which is
equal to
$v_r \partial/\partial r$
for stationary conditions,
  ${\cal C}_{brems}$
represents the bremsstrahlung cooling,
and ${\cal C}_{cyc}$ the cyclotron
cooling including the self-absorption
(Trubnikov 1958, 1973) which is
important (see expressions given
in BKL).
The rate of the energy
exchange between ions and
electrons due to the binary
collisions was obtained
by Landau (1937) and
Spitzer (1940) as
\begin{equation}
Q_{ie} \approx
{4 ~\sqrt{2\pi}~n e^4 \over m_im_e}
\left({T_e\over m_e}+
{T_i\over m_i}\right)^{-{3/2}}
\ln (\Lambda)~(T_i-T_e)~,
\end{equation}
with $\ln( \Lambda)={\cal O}(20)$
the Coulomb logarithm.
    We may express
parameters of the accretion flow as
\begin{equation}
|v_r|=\frac{\alpha ~c_s^2}{v_K {\cal J}}~, \quad
\quad
\rho=\frac{\dot M}{4 \sqrt{2}\pi \alpha }
\frac{v_K^2 {\cal J}}{r^2c_s^3}~,
\end{equation}
where ${\cal J} \equiv 1-(r_{in}/r)^{1/2}$.

   The rates of viscous heating of
ions ${\cal H}_{\eta i}$  and of electrons
${\cal H}_{\eta e}$ can be written as
\begin{equation}
{\cal H}_{\eta i}=\frac{3}{2}~
\frac{\alpha v_K c_s^2}{r}~,
\quad \quad {\cal H}_{\eta e}=
 K_e{\cal H}_{\eta i}~.
\end{equation}
 For a single temperature plasma, the
microscopic viscosity coefficient
of electrons gives $K_e =\sqrt{m_e/m_i}$
(Chapman \& Cowling 1953;
Braginskii 1965).
   In the case of  weakly interacting electrons
and ions, $K_e$ could be significantly smaller.

   Different viscosities for electrons
and ions act to give  different angular
rotation rates for electrons and ions
 and thus act to drive a toroidal
current, which is a battery effect.
  In turn this
 current  acts to
generate a poloidal magnetic field.
   However,
the high inductance of the system
 ($\propto r$) has the effect of limiting
this toroidal current to negligible values during
the time scale of accretion of the matter
(Bisnovatyi-Kogan \& Blinnikov 1977).

  In equations (12) and (13)
 the rate of Ohmic heating of ions
is written as $(1-g){\cal H}_B$ and
that of ions $g{\cal H}_B$, with
$0 \leq g \leq 1$.
  Using equations (8) and (16), we find
\begin{equation}
{{\cal H}_B \over {\cal H}_{\eta i}}
={ {{ |v_r| B^2}/ {(8 \pi  \rho )} } \over
{\alpha~v_K c_s^2 }}
= {1\over 2}~ { v_A^2 \over \alpha c_s^2}
~{|v_r| \over v_K}~.
\end{equation}
  In ~view ~of~ equation (15), this~
ratio~ is~ equal ~to
\newline $v_A^2/(2v_K^2 {\cal J})$.
  For the equipartition relation (9), this
ratio is simply $1/{\cal J}$, which
is unity except close to the inner
radius of the flow.

   The partition of the Ohmic
heating between electrons and ions,
measured by the
quantity $g$, has
a critical influence on the
model, assuming that $Q_{ie}$ is due
only to binary collisions (equation 14)
and is not altered by plasma turbulence.
   Observations of the Solar corona
outlined in \S 4 point to the
importance of magnetic field
reconnection events
and indicate that a significant part of the
magnetic field energy is dissipated by
accelerating electrons.

It follows from the physical picture of
the magnetic field reconnection
that transformation of  magnetic energy into
heat is connected with the time rate of
change of the magnetic flux and the
associated electric field in the neutral layer
which accelerates  particles.
   The electric force acting
on the electrons and the protons
is the same,
but the accelerations  are larger
by a factor $m_i/m_e$ for electrons.
   During a  short time
the electrons  gain much larger energy
than the ions.
   Additional particle
acceleration and heating
is expected at shock fronts,
appearing around
turbulent cells where reconnection occurs.
   In BKL, equations
of the form of (12) and (13)
were solved in the
approximation of nonrelativistic
electrons with the viscous and Ohmic
heating terms combined.
  Using equation (17),
the total (viscous $+$ Ohmic)
heating of electrons and of ions is
\begin{equation}
{\cal H}_i=\left(1+{2v_K^2 {\cal J}
\over v_A^2}-g\right){\cal H}_B~,
\quad \quad {\cal H}_e =g{\cal H}_B~.
\end{equation}
In the expression for
cyclotron emission, self-absorption was taken into
account using the analysis of Trubnikov (1958, 1973).
The results of calculations for
$g \sim 0.5 - 1$ show that almost
all of the energy of the electrons is radiated.
Thus the radiative efficiency of
the two-temperature, optically thin spherical
accretion flow is not less than $\sim 25\%$.
   Note  that  account the affect of
plasma turbulence on the energy
exchange rate $Q_{ie}$
could lead to a radiative
efficiency close to the value it has for
optically thick, geometrically
thin disks.

\section{Reconnection and Heating of Electrons}

  Studies of the solar corona have
led to the general conclusion
that the energy build up in
the chaotic coronal magnetic field by
slow  photospheric
driving is released by
magnetic reconnection events or flares on
a wide range of scales and energies
(Benz 1997).
    The observations support the idea of
``fast reconnection,'' not
limited to a rate proportional
to an inverse power of the
magnetic Reynolds number (Parker 1979, 1990).
   The data clearly show the
rapid acceleration of electrons (e.g.,
hard X-ray flares, Tsuneta 1996)
and ions (e.g.,
gamma ray line events and
energetic ions, Reames {\it et al.} 1997), but
the detailed mechanisms of particle
acceleration are
not established.

\begin{figure*}[t]
\epsscale{1.2}
\plotone{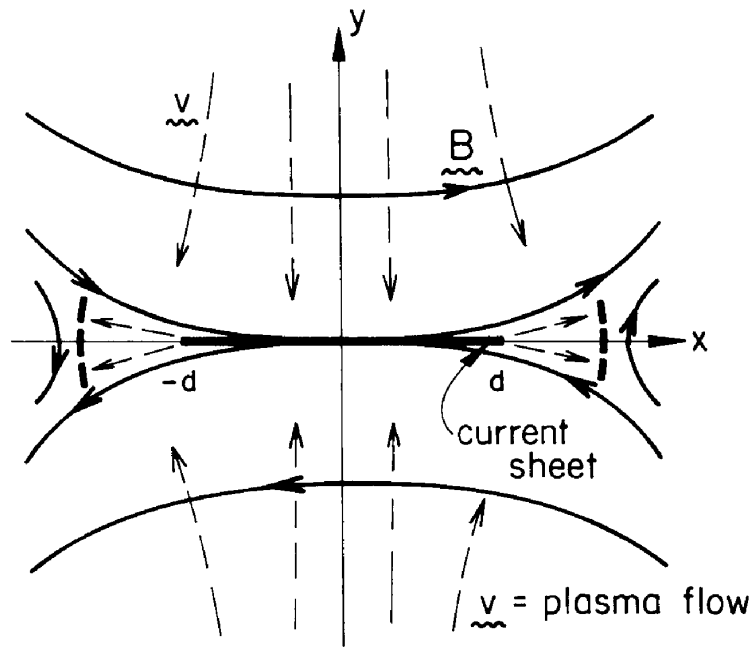}
\caption{
Sketch showing the
 magnetic
field and flow velocity
in a reconnection region
(from Romanova and Lovelace 1992).
The dashed lines indicate possible
standing shocks.
}
\label{Figure 3}
\end{figure*}

   An essential aspect of the coronal
heating is the continual input
of energy to the chaotic coronal
magnetic field due to motion of
the photospheric plasma.
    As emphasized here, there is
also a continual input of energy
to the chaotic magnetic in a quasi-spherical
accretion flow due to compression of the
flow.
   The fact that the ratio of the plasma
to magnetic pressures is small in
the corona but of order unity in the
accretion flow may of course affect
the details of the plasma processes.

  Figure 3 shows a sketch of
driven magnetic field reconnection.
  Plasma flows into the neutral
layer from above and below with
a speed of the order of the turbulent
velocity $v_t$.  Magnetic flux is
destroyed in this layer as required
for the reasons discussed earlier.
  Consequently, there
is an electric field in the $z-$direction
${\cal E}_z =-(1/c)\partial A_z/\partial t
={\cal O}( v_t B_0/c)$, where
$B_0$ is the field well outside of the
neutral layer.
  This is the same as the
estimate of the electric field given
by BKL.
   In the
vicinity of
the neutral layer ($|y| \ll \ell_t$)
this electric field
is typically
much larger than the Dreicer electric
field for electron runaway
$E_D =4\pi e^3(n_e/kT_e)\ln\Lambda$
(Parail \& Pogutse 1965), and
this  leads to streaming motion
of electrons in the $z-$direction.
   Note that it is much more
difficult to have ion runaway because
of the large gyro radii of ions
(BKL; Lesch 1991; Romanova \& Lovelace 1992).
  Of course, in the almost uniform magnetic
field assumed by Quataert (1998)
there may be no runaway of
particles.

    The streaming of the electrons
can give rise to a number of different
plasma instabilities the growth
of which gives an
anomalous resistivity.
   Galeev and
Sagdeev (1983) (see also Li {\it et al.} 1999) discuss
the quasi-linear theory
for such conditions and derive expressions
for the
rate of heating of electrons,
\begin{equation}
{d T_e \over dt} = {1\over n}
\int {d^3k \over (2\pi)^3}~
\gamma_{\bf k}~W_{\bf k} ~{{\bf k \cdot u}_{e}
\over \omega_{\bf k}}~,
\end{equation}
and ions,
\begin{equation}
{d T_i \over dt} = {1\over n}
\int {d^3k \over (2\pi)^3}~
\gamma_{\bf k}~W_{\bf k} ~,
\end{equation}
where $W_{\bf k}$ is the wavenumber energy
spectrum of the turbulence,
$\gamma_{\bf k}$
is the linear growth rate of
the mode with real frequency
$\omega_{\bf k}$, $n$ is the
number density, and ${\bf u}_e$
is the electron drift velocity.
   If the quantity
${\bf k \cdot u}_e/\omega_{\bf k}$ is
assumed constant and taken out of the
integral sign in (19), then the ratio
of heating of electrons to that of
ions is
\begin{equation}
{dT_e \over dT_i} = { u_e \over (\omega /k)}~.
\end{equation}
This result is independent of the instability
type.
    Galeev and Sagdeev (1983)  point
out that for most instabilities, relation (21)
predicts   faster heating of electrons than
ions.
    Earlier, Lesch (1991) and Di Matteo (1998)
emphasized the role of reconnection
 in accelerating electrons
to relativisitic energies in accretion
flows of AGNs.

\section{Conclusions}

    Observational evidence for a
black hole in the center of our Galaxy
and in the nuclei of other galaxies
(Cherepashchuk 1996; Haswell 1998)
make it necessary to revise or generalize
theoretical models of accretion flows.
  Improvements of the models include
account of advective terms and account
of the influence of equipartition
magnetic fields.
     Conclusions based on
ADAF solutions for optically thin
accretion flows at low mass accretion rates
are at present open to question.
    This is because
the ADAF solutions neglect
the  unavoidable magnetic
field annihilation which gives significant
electron heating.
   In contrast with the ADAF solutions,
we show that the field
annihilation leads to a radiative
efficiency $\gtrsim 25\%$ of the
standard value for  an optically
thick, geometrically thin disk.
   It is possible that a full
treatement of the ion-electron energy
exhange due to the plasma turbulence
further increases the radiative efficiency to
this standard value  (see also
Fabian \& Rees 1995).

  Some observational data which
were interpreted as evidence for the
existence of the ADAF regime
have disappeared after additional
accumulation of data.
  The most interesting example
of this sort is connected with the
claim of  ``proof'' of the
existence of event horizons
of  black holes
due to manifestation of the
ADAF regime of accretion
(Narayan {\it et al.} 1997).
   Analysis of a more complete
set of  observational data
(Chen {\it et al.} 1997)  shows
that the statistical effect
claimed as an evidence for ADAF disappears.
    This example shows the danger of
``proving'' a  theoretical
model with preliminary observational data.
   It is even more dangerous
when the model is physically
not fully consistent because
then even a reliable set
of the observational
data cannot serve as a proof
of the model.
   The classical example from
astrophysics of this kind
gives the theory of the
origin of the elements
presented in the famous book
of Gamow (1952), where the model of
the hot universe was developed.
   In addition to the remarkable
predictions of this model,
the author  wanted to explain
the origin of heavy elements in the primordial
explosion, neglecting the
problems connected with the absence of
stable elements with the
number of baryons equal to $5$ and $8$.
  Gamow found good agreement with
his calculations, where the mentioned
problem was neglected, and considered
the observational curve as a proof of
his theory of the origin of the elements.
  Further developments
have shown that his outstanding
theory explains many things, except
the origin of the heavy elements,
which are produced as a result of stellar
evolution.

     Thus, there fundamental reasons
for questioning the application
ADAF models to
 under-luminous AGN, where the
observed energy flux is much smaller
than  expected for
standard accretion disk models.
   Two possible explanations
 may be suggested.
One is based on a more
accurate estimations
 of the accretion mass flow
into the black hole, which could be
 overestimated.
    Another possibility is based
 on existence of jets and/or
uncollimated outflows which carry
away most of the accreting matter.
    Many compact astrophysical objects
thought to have  accretion disks
are observed to have jets or outflows,
including
active galaxies and
quasars,  old
compact stars in binaries,
 and  young stellar
objects.
    The formation of jets and outflows
is very probable under conditions where
 an ordered magnetic field threads a
thin disk (see reviews by
Bisnovatyi-Kogan 1993;  Lovelace,
Ustyugova, \& Koldoba 1999).
  To extend this interpretation, we suggest that
under-luminous AGNs may lose the main
part of their energy to the formation of jets or
outflows.
   This suggests a search for a correlation
between existence of jets or
outflows and under-luminous
galactic nuclei.

\acknowledgements
{We thank M.M. Romanova for
valuable discussions.
   This work was supported in part by
NSF grant AST-9320068 and NASA
grant NAG5 6311.
    Also, this work
was made possible in part by
Grant No. RP1-173 of the U.S.
Civilian R\&D Foundation for the
Independent States of the
Former Soviet Union.
   The work of
GBK was also supported by Russian
Fundamental Research Foundation
grant No. 96-02-16553.  }


\begin{thebibliography}{}




\bibitem{} Benz A.O. 1997, in {\it Solar
and Heliospheric Plasma Physics}, G.M.
Simnett, C.E. Alissandrakis, \& L. Vlahos,
eds. (Heidelberg: Springer), 201


\bibitem{} Bisnovatyi-Kogan, G.S. 1993,
in {\it Stellar Jets and Bipolar
Outflows,} L. Errico \& A.A. Vittone, eds.
(Dordrecht: Kluwer), 369

\bibitem{} Bisnovatyi-Kogan, G.S., \&
Blinnikov, S.I. 1977, A\&A, 59, 111

\bibitem{}
Bisnovatyi-Kogan, G.S. \& Lovelace, R.V.E.
1997, ApJ, 486, L43

\bibitem{}
Bisnovatyi-Kogan, G.S., \& Ruzmaikin, A.A., 1974,
Astrophys. and Space Sci., 28, 45

\bibitem{}
Bisnovatyi-Kogan, G.S., \& Ruzmaikin, A.A., 1976,
Astrophys. and Space Sci., 42, 401

\bibitem{}
Bisnovatyi-Kogan, G.S. and Sunyaev, R.A.
1972, Soviet Astron., 15, 697

\bibitem{}
Blackman, E.G.
1998, \mnras, in press (astro-ph/9710137)

\bibitem{} Braginskii, S.I. 1965, in {\it Reviews
of Plasma Physics}, Vol. 1, M.A. Leontovich, Ed.
(Consultants Bureau: New York), p. 205

\bibitem{} Cannizzo, J.K., Greenhill, L.J.,
Herrnstein, J.R., Moran, J.M., \& Mushotsky, R.R.
1998, Am. Astron. Soc. Meeting No. 192, abstract 41.03

\bibitem{}
Chang, K.M, \& Ostriker, J.P.
1985, \apj, 288, 428

\bibitem{}
Chapman, S., \& Cowling, T.G. 1953,
{\it The Mathematical Theory of Non-Uniform
Gases} (Cambridge Univ. Press)

\bibitem{}
Chen, W., Cui, W., Frank, J., King, A.,
Livio, M., \& Zhang, S.N.
1997, Talk at High Energy Astrophysics
Division Meeting, November, 1997

\bibitem{}
Cherepashchuk, A.M.
1996, Uspekhi Fiz. Nauk, 166, 809

\bibitem{} Di Matteo, T., 1998, \mnras, 299, L15

\bibitem{} Di Matteo, T., Fabian, A.C., Rees, M.J.,
Carilli, C.L., \& Ivison, R.J., 1998, \mnras,
in press (astro-ph/9807245)


\bibitem{}
Fabian, A.C. and Rees, M.J.
1995, Month. Not. R.A.S., 277, L55

\bibitem{} Galeev, A.A., \& Sagdeev, R.Z. 1983,
Chap. 6, in {\it Handbook of Plasma Physics}, Vol. 1,
eds. Rosenbluth, M.N., \& Sagdeev, R.Z.
(Amsterdam: North-Holland Pub.), Ch. 6.1

\bibitem{}
Gamow, G. 1952,
{\it The Creation of the Universe},
(Viking Press: New York)

\bibitem{} Goldreich, P., \& Sridhar, S.
1995, \apj, 438, 763

\bibitem{}
Herrnstein, J.R., Moran, J.M., Greenhill, L.J.,
Blackman, E.G., \& Diamond, P.J. 1998, \apj, 508, 243



\bibitem{}
Ho, L. 1999, in {\it Observational Evidence
for Black Holes in the Universe},
ed. S. Chakrabarty,
(Kluwer), in press

\bibitem{} Ichimaru, S. 1977, \apj, 214, 840

\bibitem{}
Landau, L.D. 1937,
Zh. Exp. Theor. Phys. 7, 203.

\bibitem{}
Lasota, J.-P., Abramovicz M.A., Chen, X.,
Krolik, J., \& Narayan, R., Yi, I.
1996, ApJ, 462, 142

\bibitem{} Lesch, H. 1991, A\&A, 245, 48

\bibitem{} Li, H., Colgate, S.A., Kusunose, M.,
\& Lovelace, R.V.E. 1999,
to appear in
{\it Proc. of High Energy Processes in Accreting
Black Holes}, eds.,
J.Poutanen, \&  R. Svensson  (ASP Conf. Series)


\bibitem{} Lovelace, R.V.E., Ustyugova, G.V.,
\& Koldoba, A.V. 1999, to appear in {\it
Proc. of IAU Symposium 194}, in press
(astro-ph/9901256)



\bibitem{}
Menou, K., Quataert, E., \& Narayan, R.
1997, to appear in {\it Proc. of 8th Marcel
Grossman Meeting on General Relativity},
in press (astro-ph/9712015)

\bibitem{}
Narayan, R., Barret, D., \& McClintock, J.E.
1997, \apj, 482, 448


\bibitem{}
Narayan, R., Garcia, M.R., \& McClintock, J.E.
1997, \apj, 478, L79



\bibitem{} Narayan, R., Mahadevan, R., \&
Quataert, E. 1999, to appear in {\it The
Theory of Black Hole Accretion Discs},
eds. Abramowicz, M.A., Bjornsson, \& Pringle,
J.E. (Cambridge University Press)
(astro-ph/9803141)

\bibitem{} Narayan, R., \& Yi, I.
1995, ApJ, 452, 710

\bibitem{} Parail, V.V., \& Pogutse, O.P. 1965,
in {\it Reviews of Plasma Physics}, V. 11,
ed. M.A. Leontovich (New York: Consultants
Bureau), 1


\bibitem{} Parker, E.N. 1979, {\it Cosmical
Magnetic Fields}, (Clarendon, Oxford)


\bibitem{} Parker, E.N. 1990, in
{\it Mechanisms of Chromospheric and
Coronal Heating}, eds. P. Ulmschneider,
E.R. Priest, \& R. Rosner (Berlin:
Springer-Verlag), 615


\bibitem{}
Quataert, E.
1998, \apj, 500, 978

\bibitem{} Reames, D.V., Barbier, L.M.,
Von Rosenvinge, T.T., Mason, G.M., Mazur, J.E.,
\& Dwyer, J.R. 1997, \apj, 483, 515

\bibitem{} Romanova, M.M.,
\& Lovelace, R.V.E. 1992, A\&A, 262, 26

\bibitem{}
Shvartsman, V.F.
1971, Soviet Astron., 15, 377

\bibitem{}
Shakura, N.I. 1972, Astron. Zh., 49, 921
(1973, Sov. Astron., 16, 756)

\bibitem{} Shakura, N.I.,
\& Sunyaev, R.A. 1973, A\&A,
24, 337


\bibitem{}
Shapiro, S.L., Lightman, A.P., \& Eardley, D.M.
1976, ApJ, 204, 187

\bibitem{}
Spitzer, L.
1940, MNRAS, 100, 396

\bibitem{} Trubnikov, B.A. 1958,
{\it Magnetic Emission of
High Temperature Plasma,}
Dissertation, Moscow, USAEC Tech.
Information Service, AEC-tr-4073 (1960)


\bibitem{}
Trubnikov B.A.
1973, Voprosy Teorii Plasmy, 7, 274

\bibitem{ts96}
Tsuneta, S.
1996, ApJ, 456, 840.

\end{thebibliography}
\end{document}